\documentstyle[12pt,openbib,epsf]{article}
\newcommand{\postscript}[2]
{\setlength{\epsfxsize}{#2\hsize}
\centerline{\epsfbox{#1}}}
\begin{document}
\begin{titlepage}
\begin{center}
{\bf Shape-Independent
Expansion for the $^3S_1-^3D_1$ Mixing Parameter}\\
\vskip 1.5cm

{Sadhan K. Adhikari,$^{a}$ Lauro Tomio,$^{a}$ J. P. B. C. de
Melo,$^{b}$ and T. Frederico$^{c}$ \\
\vskip 0.5cm
$^{a}$Instituto de F\'\i sica Te\'orica, Universidade Estadual Paulista\\
01405-001 S\~{a}o Paulo, S\~{a}o Paulo, Brasil\\
$^{b}$Instituto de F\'\i sica, Universidade de S\~ao Paulo,
01498-970 S\~ao Paulo, S\~ao Paulo, Brasil\\
$^{c}$Instituto de Estudos Avan\c cados, Centro T\'ecnico Aeroespacial\\
12231-970 S\~ao Jos\'e dos Campos,  S\~{a}o Paulo, Brasil
}
\end{center}
\vskip 1cm
\begin{abstract}
A \ low - energy \ shape - independent \ expansion \ is \ suggested \
for \ the \ function \ $[tan(2\epsilon_{BB})/(2k^2)]$, \ where $\epsilon_{BB}$
i
   s the
Blatt-Biedenharn mixing parameter for the $^3S_1 -$ $^3D_1$ channel. This
expansion allows an evaluation of the mixing parameter $\epsilon_{BB}$
 from a knowledge of the deuteron asymptotic $D$ to $S$ ratio, pion
mass and other low-energy  observables, such as the scattering lengths,
deuteron binding etc., of the nucleon-nucleon  system.
We demonstrate that the correct long range
behavior of the tensor potential is essential for a realistic reproduction of
$\epsilon_{BB}$.
\end{abstract}
\vskip 0.5cm
\end{titlepage}
The following effective-range expansion has been extremely useful for
low-energy phase-shift analysis\cite{bethe,weiss}
\begin{equation}
k^{2l+1} cot \delta_l =-\frac{1}{a}+\frac{1}{2}r_0 k^2 +...,
\label{eff} \end{equation}
where $a$ is the scattering length, $r_0$ is the effective range, $k^2$ is the
c.m. energy and $\delta_l$ is the phase-shift.  We employ units
$\hbar=m_n=1$, where $m_n$ is the nucleon mass. The parameters $a$ and $r_0$
are determined by some average properties of the potential and are independent
 of detailed shape of the potential. This is why these expansions are termed
 shape independent. In the coupled $^3S_1-^3D_1$ channel, apart from the phase
 shifts, one needs the mixing parameter for a complete analysis. Here we
 provide an effective-range-type shape independent expansion for the mixing
 parameter.

There are two commonly used mixing parameters, the
Stapp-Ypsilantis\cite{stapp} mixing parameter $\epsilon_1$ and the
Blatt-Biedenharn\cite{blatt1} mixing parameter $\epsilon_{BB}$. In the
present treatment we shall only consider the Blatt-Biedenharn mixing parameter
as it is convenient to define an analytic function appropriate for making
the desired expansion involving this  parameter. This expansion involves the
deuteron $D$ state to $S$ state ratio, $\eta_d$, pion mass, $m_{\pi}$, and
other on-shell quantities such as, the nucleon-nucleon ($NN$) binding energies
and scattering lengths and allows one to evaluate the mixing parameter
$\epsilon_{BB}$ at low energies in a model independent way.
Lacking precise measurements for $\epsilon_{BB}$, we illustrate the present
approach using the theoretically calculated mixing parameters of model
NN potentials.

It is interesting to point out that there have  been various studies of
model-independent correlations involving  $D$ state observables
of the $NN$ system. For example, there have been studies of correlations
 among $\eta_d$ and the deuteron quadrupole moment
$Q$,\cite{erics} and among $\epsilon$ and $Q$.\cite{musta}

Ever since Amado and coworkers\cite{amado1} pointed out the importance of
  $\eta_d$, in measuring the strength of the $D$ state, there has
 been lot of theoretical and experimental activities\cite{erics,acti1}
in measuring $\eta_d$. The present expansion could be used for a
 evaluation of $\eta_d$  once accurate low-energy  mixing parameters
  $\epsilon_{BB}$ are available.

It is convenient to consider  the following
`effective-range-type' function appropriate for making expansion\cite{blatt1}
\begin{equation}
\frac{tan(2\epsilon_{BB})}{2k^2}\equiv \frac {\hat K_{02}/k^2}{\hat K_{00}-\hat
K_{22}},
\label{50} \end{equation}
which is analytic in $k^2$ in the scattering region. We
use notation $\hat K_{ll'}\equiv K_{ll'}(k^2)$, where
$K_{ll'}(k^2) \equiv K_{ll'}(k,k;k^2)$ is the on-shell $NN$
$K$ matrix element for the $ll'$ channel.
This function  satisfies the two following limits:
\begin{equation}
\lim_{k\to 0}\frac{tan(2\epsilon_{BB})}{2k^2}= \lim_{k\to 0}
\frac {\hat K_{02}/k^2}{\hat K_{00}-\hat K_{22}}\equiv\frac{a_{02}}{a_t},
\label{60} \end{equation}
and
\begin{equation}
\lim_{k\to i\alpha}\frac{tan(2\epsilon_{BB})}{2k^2}= \lim_{k\to i\alpha}
\frac {\hat K_{02}/k^2}{\hat K_{00}-\hat K_{22}}\simeq
\frac{\eta_{d}}{\alpha^2},
\label{70} \end{equation}
where $a_{t}$ is the  triplet scattering length for the (00) channel in units
of fm, $a_{02}$ is the scattering length for the (02) channel in
units of fm$^3$  defined by
$\lim_{k\to 0}\hat K_{02}/k^2 \equiv a_{02}$ and  $\alpha^2$ is the deuteron
binding energy in units of fm$^{-2}$. At the deuteron bound-state pole the
approximation, $\lim_{k\to i\alpha} \hat K_{02}/[\hat K_{00}-\hat K_{22}]
=-\eta_d$, holds within an  estimated  error  of less than
0.1\%, in place of the
exact relation  $\lim_{k\to i\alpha} \hat K_{02}/\hat K_{00}=-\eta_d$, so
for all practical purposes Eq. (\ref{70}) is taken to be exact.
Both $\eta_d$ and $a_{02}$ are to be taken as  measures of the strength of
the tensor force.

In realistic situations
the two limits of the function $tan(2\epsilon_{BB})
/(2k^2)$ given by Eqs. (\ref{60}) and (\ref{70}) are quite different and this
is a consequence of including a realistic long range behavior
of the $NN$ interaction. These two limits are essential for a correct
reproduction of the mixing parameters. In the
case of the Yamaguchi\cite{yamag} tensor potential these two limits are
practically the same because of the slowly varying form factors of this
potential. This is why the Yamaguchi tensor potential usually fails to
reproduce even qualitatively the mixing parameters.

We consider the following function $B(k^2)$  for our purpose
\begin{equation}
B(k^2) \equiv -\frac {\eta_d}{\alpha^2} +
\frac{\hat K_{02}/k^2 }{\hat K_{00}-\hat K_{22}}.
\label{1} \end{equation}
The $K$ matrix elements are analytic functions of $k^2$ for positive energies
and have the left-hand potential cuts. Hence the
 function $B(k^2)$ also  has similar analytic properties. Particularly,
it is analytic in the whole complex energy plane except
the left-hand meson-exchange cuts  at negative
energies for $-\infty < k^2 < -m_{\pi}^2/4$, where $m_{\pi}$ is the pion mass.
Equations (\ref{60}) and (\ref{70}) imply
\begin{equation}
B(-\alpha ^2)=0
\label{3} \end{equation}
and
\begin{equation}
B(0)=\frac{a_{02}}{a_t}-\frac{\eta_d}{\alpha^2}. \label{110} \end{equation}

Using the above properties we  suggest a convenient way of parametrizing
the function $B(k^2)$.
The left-hand cut of this function is essential for a correct reproduction
of the mixing parameters.
It is well-known that in region remote from the cuts the
function $B(k^2)$ could be parametrized by approximating the nearest
branch-point by a pole, e.g., at $k^2 = -m_{\pi}^2/4$, which corresponds to
the exchange of a pion. Recalling, also, that
$B(-\alpha ^2) = 0$, one has  the following Taylor-series-type expansion
\begin{equation}
B(k^2) = \frac{k^2+\alpha^2}{k^2+m_{\pi}^2/4} [c_1+c_2(k^2+\alpha^2)+...],
\label{AX} \end{equation}
where $c_1$, $c_2$ etc. are parameters
essentially determined by some average properties of the tensor force and
low-energy on-shell  $NN$ observables. For the usual effective-range
function (\ref{eff}) the effect of the left-hand cut is less prominent.
Hence simple $NN$ potentials, such as, Yamaguchi and square-well potentials,
provide a good description of the phase shifts, but not the mixing
parameters.

At low energies, the parameter
$c_1$ alone is supposed to contribute to $B(k^2)$ and $tan(2\epsilon_{BB})$.
Consequently, setting $k=0$, keeping only the term involving $c_1$
in Eq. (\ref{AX}) and using Eq. (\ref{110}),
one has the following approximate relation
\begin{equation} c_1 \simeq \frac {m_\pi^2}{4\alpha^2}B(0)
=\frac{m_{\pi}^2}{4\alpha^2}\left(
\frac{a_{02}}{a_t}-\frac{\eta_{d}}{\alpha^2}\right), \label{80}
\end{equation}
and $c_2$ = 0.
At higher energies, however,  both $c_1$ and $c_2$ are supposed to
contribute.

{}From  Eqs. (\ref{50}), (\ref{1}) and (\ref{AX}) we obtain the following
 expansion for the  parameter $\epsilon_{BB}$
\begin{equation}
\frac{tan(2\epsilon_{BB})}{2k^2}  = \frac{\eta_d}{\alpha^2}+
\frac{k^2+\alpha^2}{k^2+m_{\pi}^2/4} [c_1+c_2(k^2+\alpha^2)+...],
\label{90}
\end{equation}
which reduces at low energies to the following shape independent
approximation
\begin{equation}
\frac{tan(2\epsilon_{BB})}{2k^2}  = \frac{\eta_d}{\alpha^2}+
\frac{m_\pi^2}{4\alpha^2}\left( \frac{a_{02}}{a_t}-\frac{\eta_d}{\alpha^2}
\right) \frac{k^2+\alpha^2}{k^2+m_\pi^2/4} .
\label{120}
\end{equation}
Expansions (\ref{90}) and (\ref{120}) are the desired expansions. In Eq.
(\ref{120}) all the quantities are zero- or negative-energy on-shell
observables. These observables are determined by some average properties
of the potential and not by its full intricacy. The new on-shell parameter
of interest in this case is $a_{02}$. Alternatively, this parameter
 is related to  $c_1$ and $B(0)$ of Eq. (\ref{80}).

We use expansions  (\ref{90}) and (\ref{120}) for predicting
the  mixing parameters $\epsilon_{BB}$.
In the absence of accurate experimental mixing parameters\cite{expt}
 we illustrate the present approach using
the numerical results  of  the Peiper modified Reid soft core
(PRSC)\cite{peiper} and the momentum space
one boson exchange Bonn (OBE)\cite{bonn1}
potentials for the $^3S_1-^3D_1$ channel. For the PRSC potential,
deuteron binding is 2.2298 MeV, $\eta_d = 0.02636$, $a_t=5.386$ fm, and
$a_{02}= 1.62$ fm$^3$. The pion mass $m_\pi$ is taken to be 0.7 fm$^{-1}$.
Using these parameters in Eq. (\ref{80}) we
obtain $B(0)$=$-$0.190 fm$^2$ and $c_1=-0.432$ fm$^2$.
 For the OBE  potential\cite{bonn1} deuteron binding is 2.2245 MeV,
 $a_t=5.4218$ fm, $\eta_d=0.0267$, $a_{02}=1.71$ fm$^3$. Using
 these parameters in Eq. (\ref{80}) we
obtain $B(0)$=$-$0.182 fm$^2$ and $c_1=-0.416$ fm$^2$. If we use these
theoretical parameters in Eq. (\ref{120}) we have an excellent fit
to the numerically calculated $\epsilon_{BB}$ for both these potentials at low
energies: $E_{lab} < 4$ MeV.
However, at higher energies ($E_{lab} < 100$ MeV) we need to
modify the numerical value of these parameters in order to obtain a good fit.
In Fig. 1 we plot the numerically calculated mixing parameters
 $\epsilon_{BB}$
(${\large{\diamond}} -$ PRSC; $\times - $ OBE) and the present fit via Eq.
(\ref{120}) (full line $-$ PRSC; broken line $-$ OBE)
versus $E_{lab}$. For PRSC (OBE) potentials we used
$B(0)$ = $-$0.191 fm$^2$    ($-$0.180 fm$^2$),
$c_1$ = $-$0.458 fm$^2$ (=$-$0.482 fm$^2$).  The small deviation of $c_1$ at
higher energies from the theoretically calculated value
is due to the approximate nature of Eq. (\ref{80}), which is strictly
valid for low energies.  For still higher energies
($E_{lab} < 300$ MeV) we need both $c_1$ and $c_2$ for a correct reproduction
of $\epsilon_{BB}$. In Fig. 2 we exhibit the numerically calculated
 $\epsilon_{BB}$ for  PRSC (OBE) potentials  and the present fit via Eq.
 (\ref{90}) with
$c_1$ = $-$0.468 fm$^2$ (=$-$0.492 fm$^2$), and
$c_2$ = 0.014 fm$^4$ (= 0.008 fm$^4$).
The agreement between the theoretical mixing parameters and those
obtained with the shape independent expansion is good in all cases.

The value of $B(0)$ of Eq. (\ref{110})
is particularly interesting because it essentially
determines the low-energy behavior of $tan(2\epsilon_{BB})/2$ given by
Eq. (\ref{120}). We  provide an approximation
to $B(0)$ which we use to show that this parameter is
essentially determined by the long-range part of the tensor force and
 on-shell $S$ wave $NN$ observables.
The present approximation to $B(0)$ could be obtained by
dividing the explicit  on-shell Lippmann-Schwinger
equation for $K_{02}$ by $(k^2\hat K_{00})$:
\begin{eqnarray}
\frac{\hat K_{02} }{ \hat K_{00} k^2}
 =  \frac{  V_{02}(k,k) }{ \hat K_{00}k^2 }
+ \frac{2}{\pi}{\cal {P}}\int_0^\infty  \frac{dq q^2  K_{00}(k,q;k^2)
V_{02}(q,k)}
{\hat K_{00}k^2(k^2-q^2)}+ \frac{<k \mid  K_{02} G_0 V_{22}
 \mid k>}{\hat K_{00}k^2},
\label{5} \end{eqnarray}
where ${\cal {P}}$ denotes the principal value prescription.
Subtracting Eq. (\ref{5}) at $k=i\alpha$ from Eq. (\ref{5}) at $k=0$  one gets
the following approximate relation
\begin{eqnarray}
B(0)  \simeq \frac{1}{a_t} \lim_{k\to 0}\left(\frac{V_{02} (k,k)}
{k^2}\right)-\frac{2}{\pi}\alpha^2\int_0^\infty dq  \frac{K_{00}(0,q;0) }
{a_t(\alpha^2+q^2)}\lim_{k\to 0}\left(\frac{V_{02}(q,k)}{k^2}\right).
\label{10} \end{eqnarray}
In deriving Eq. (\ref{10}) we have ignored the difference between the
higher order terms of the scattering equation (\ref{5})
involving $V_{22}$ and made approximations, such as
$ lim_{k\to 0}V_{02}(q,k)/k^2 \simeq  -V_{02}(q,i\alpha)/\alpha^2$.

In Eq. (\ref{10}) the low-$q$ values of the tensor potential $V_{02}(q,k)$,
and the $K$ matrix elements  dominate the  integral because
of additional factors of $q^2$ in the denominator for large $q$.
 It is well known that the half-shell function $g(q)\equiv K(0,q;0)/a_t$ is a
universal function independent of potential models.\cite{redish} We find that
any reasonable approximation to this quantity
 in Eq. (\ref{10})  leads
 to the same result to less than  an estimated error of 1 $\%$.
 We calculated $B(0)$ with the PRSC
 $NN$ tensor potentials $V_{02}$\cite{peiper};
 for $g(q)$   we used the exact results
 for the PRSC potential and  the
 Yamaguchi potential: $g(q)=\beta^2/(q^2+\beta^2), \beta=1.4$ fm$^{-1}$.
 We obtained from Eq. (\ref{10}) $B(0)=-0.184$ fm$^2$,
 using PRSC half shell functions; and $B(0)=-0.183   $ fm$^2$,
 with the Yamaguchi form-factor. This shows that Eq. (\ref{10}) provides
 a very good approximation to the exact $B(0)$, given by Eq. (\ref{110}),
  which in this  case  is $-$0.190 fm$^2$.

  The approximation (\ref{10}) to $B(0)$ can also
 be used to demonstrate the importance of the long-range behavior of
 the tensor potential. The PRSC tensor potential\cite{peiper}
 $V_{02}$ is a superposition of four Yukawa type potentials, with the longest
 range component simulating the exchange of a
 pion with a range parameter of $\mu$ = 0.7 fm$^{-1}$. The three other
 components of this potential simulates exchange of mesons
 of masses $2m_\pi$, $4m_\pi$, and $6m_\pi$. We   evaluated
  $B(0)$ using Eq.  (\ref{10}) and   setting the strengths of the components
of the PRSC tensor potential $V_{02}$ corresponding to exchanged mesons of
masses $4m_\pi$, and $6m_\pi$ to zero, using both Yamaguchi and PRSC half
shell functions. We obtained for $B(0)$ essentially the previous value
$B(0) = -0.184$ fm$^2$ in both cases. This shows that
$B(0)$ is  insensitive to the short range part of the tensor potential.
Finally, we calculated $B(0)$  with only the longest range part of tensor
PRSC potential $V_{02}$ corresponding to the exchanged
pion. With both types of half-shell functions we obtained in this case
$B(0) = -0.162$ fm$^2$. This corresponds to a percentage error of only 12$\%$
in relation to the full potential: $B(0) = -0.184$ fm$^2$. Hence only the
correct pion exchange tail of the potential approximately reproduces $B(0)$.
The Yamaguchi tensor potential, which does not have this long range
behavior yields for $B(0)$ of Eq. (\ref{110}) the very small value
$B(0) = 0.02$ fm$^2$, and fails to reproduce even approximately the
realistic mixing parameters.  This simple calculation demonstrates
the importance of the correct long-range behavior
 of the tensor potential $V_{02}$
in reproducing the exact mixing parameters.

In short, we have presented an effective-range-type expansion  for
 the on-shell quantity $tan(2\epsilon_{BB})/(2k^2) \equiv$
$(\hat K_{02}/k^2)/{[\hat K_{00}-\hat K_{22}]}$ valid at low energies, which
can be used for  predicting
the mixing parameter $\epsilon_{BB}$ using the experimental
$\eta_d$, deuteron binding, $NN$ scattering lengths $a_t$ and $a_{02}$,
 pion mass $m_\pi$. The parameter $a_{02}$ (or $B(0)$) is essentially
 determined by the correct long-range behavior
  of the $NN$ tensor potential, which is demonstrated
 to be essential for a realistic prediction of the mixing parameter
 $\epsilon_{BB}$.

We thank Dr. G. Krein for his valuable helps.
The work has been supported in part by the Conselho Nacional de
Desenvolvimento $-$ Cient\'\i fico e Tecnol\'ogico of Brasil.

%
\begin{figure}[p]
\postscript{eta1.ps}{1.0}
\vskip -10cm
\caption[]
{The numerically calculated Blatt-Biedenharn mixing parameters in
radians for the PRSC\cite{peiper} (${\large{\diamond}}$) and the
 OBE\cite{bonn1} ($\times$) potentials and the present fit,
with $c_1 = -0.458$ fm$^2$ (PRSC $-$ full line), $=-0.482$ fm$^2$
(OBE $-$ broken line) and $c_2 = 0$, for $E_{lab}$ upto 100 MeV. }
\label{fig. 1}
\end{figure}
\begin{figure}[p]
\vskip -10cm
\postscript{eta2.ps}{1.0}
\caption[]
{Same as Fig. 1
with $c_1 = -0.468$ fm$^2$, $c_2 = 0.014$ fm$^4$ (PRSC $-$ full line),
$c_1=-0.492$ fm$^2$ and $c_2=0.008$ fm$^4$
(OBE $-$ broken line), for $E_{lab}$ upto 300 MeV.}
\label{fig. 2}
\end{figure}
{\bf {Figure Caption}}
\vskip .5cm
\noindent {\bf 1.} The numerically calculated Blatt-Biedenharn
mixing parameters in radians for the PRSC\cite{peiper}
(${\large{\diamond}}$) and the OBE\cite{bonn1} ($\times$)
potentials and the present fit via Eq. (\ref{90}) with $c_1 =
-0.458$ fm$^2$ (PRSC $-$ full line), $=-0.482$ fm$^2$ (OBE $-$
broken line) and $c_2 = 0$, for $E_{lab}$ upto 100 MeV.
\vskip .5cm
\noindent {\bf 2.} Same as Fig. 1
with $c_1 = -0.468$ fm$^2$, $c_2 = 0.014$ fm$^4$ (PRSC $-$ full line),
$c_1=-0.492$ fm$^2$ and $c_2=0.008$ fm$^4$
(OBE $-$ broken line), for $E_{lab}$ upto 300 MeV.

\begin{thebibliography}{99}
\bibitem{bethe} H. A. Bethe, Phys. Rev. 76 (1949) 38.
\bibitem{weiss} J. M. Blatt and V. F. Weisskopf, Theoretical Nuclear
Physics, (Wiley, New York, 1952).
\bibitem{stapp} H. P. Stapp, T. Ypsilantis, and N. Metropoulis, Phys. Rev.
105 (1957) 302.
\bibitem{blatt1} J. M. Blatt and L. C. Biedenharn, Phys. Rev. 86 (1952) 399;
 L. C. Biedenharn and J. M. Blatt, ibid. 93
(1954) 1387.
\bibitem{erics} T. E. O. Ericson and M. P. Rosa Clot, Ann. Rev. Nucl. Part.
Sci. 35 (1985) 271.
\bibitem{musta} M. M. Mustafa, Phys. Rev. C 47 (1993) 473;
D. W. L. Sprung, H. Wu, and J. Martorell, ibid. 42 (1990) 863;
W. van Dijk, M. W. Kermode, D.-C. Zheng, ibid. 47 (1993) 1898.
\bibitem{amado1} R. D. Amado, Phys. Rev. C 19 (1974) 1473;
 R. D. Amado, M. P. Locher, and M. Simonius, Phys. Rev. C 17 (1978) 403.
\bibitem{acti1}J. T. Londergan, C. E. Price, and E. J. Stephenson, Phys.
Rev. C 35 (1987) 902;
L. D. Knutson, P. C. Colby, and B. P. Hichwa, Phys. Rev. C 24 (1981) 411;
L. D. Knutson, B. P. Hichwa, A. Barrosso, A. M. Eiro, F. D. Santos, and
R. C. Johnson, Phys. Rev. Lett. 35 (1975) 1570.
\bibitem{yamag} Y. Yamaguchi and Y. Yamaguchi, Phys. Rev. 95 (1954) 1635.
\bibitem{expt} A. R. Arndt, I. D. Roper, R. A. Bryan, R. B. Clark, B. J.
VerWest, and P. Signell, Phys. Rev. D 28 (1983) 97.
\bibitem{peiper} S. C. Peiper, Phys. Rev. C 9 (1974) 883; R. V. Reid,
Ann. Phys. (N.Y.) 50 (1968) 411.
\bibitem{bonn1} R. Machleidt, K. Holinde, and Ch. Elster, Phys. Rep.
149 (1987) 1.
\bibitem{redish} E. F. Redish and K. Stricker Bauer, Phys. Rev. C. 36
 (1987) 513; K. Amos et al., ibid. 37 (1988) 934.

\end{thebibliography}
\end{document}